\documentclass[12pt]{article}
\input epsf.sty

\def\lromn#1{\uppercase\expandafter{\romannumeral#1}}

\usepackage{graphicx,amsmath,amssymb}
\begin{document}

\begin{center}
\begin{large}
\textbf{
Solitons and Precision Neutrino Mass Spectroscopy
}

\end{large}
\end{center}

\vspace{2cm}
\begin{center}
\begin{large}
M. Yoshimura

Center of Quantum Universe, Faculty of
Science, Okayama University \\
Tsushima-naka 3-1-1 Kita-ku Okayama
700-8530 Japan
\end{large}
\end{center}

\vspace{5cm}

\begin{center}
\begin{Large}
{\bf ABSTRACT}
\end{Large}
\end{center}

We propose how to implement  precision
neutrino mass spectroscopy
using radiative neutrino pair emission (RNPE)
from a macro-coherent decay of a
new form of  target state, 
a large number of activated atoms interacting with static condensate field.
This method makes it possible to
measure still undetermined parameters 
of the neutrino mass matrix, two CP violating Majorana phases, 
the unknown mixing angle and
the smallest neutrino mass which could be of order a few meV,
determining at the same time the Majorana or Dirac nature of masses.
The twin process of paired superradiance (PSR) is also discussed.

\newpage
{\bf Introduction}
\hspace{0.2cm}
Neutrinos are still mysterious particles:
their absolute mass scale (or the smallest neutrino mass), the nature of masses
(whether they have Majorana or Dirac type masses),
and their relation to the leptogenesis theory
\cite{fy-86}, \cite{davidson-ibarra}
are not clarified experimentally.
Experimental efforts to unravel these properties are
mainly focused on nuclear targets.
Nuclear targets, however, are problematic
at least in one important aspect, the mismatch of energy scale:
the released energy of nuclear transition is of order several MeV,
and this is far separated from the expected
neutrino mass range of O[0.1]eV.

We proposed a few years ago the idea of using atomic targets
to overcome this difficulty; RNPE
from a metastable state $|e\rangle$,
$|e\rangle \rightarrow | g\rangle + \gamma+ \nu_i  \nu_j$.
This is an elementary process predicted by the ordinary electroweak
interaction, and its detection opens a path towards
the neutrino mass spectroscopy
\cite{my-06}, \cite{pv},
by precisely measuring the  photon energy spectrum,
thereby resolving neutrino mass eigenstates
$\nu_i\,, i=1,2,3$.

With smaller released energies of atomic
transitions, the atomic decay
involving neutrino pair emission
has a demerit of tiny weak rates,
unless  a new idea of  rate enhancement is taken into account.
Our enhancement mechanism uses a coherent cooperative effect of a large
number of atoms interacting with a common field 
\cite{macro-coherence}, \cite{my-10-10}.
A similar idea goes back to the superradiance (SR for short) 
 \cite{sr review} of a single photon emission,
where the decay rate from many atoms
is in proportion to $n^2V$,
the target number density squared times a coherent volume $V$,
unlike the target number $nV$ in the spontaneous decay.

Atoms in a metastable state $|e\rangle$ may have a lifetime
for a long time measurement.
If these atoms  further have a developed coherence,
macro-coherent two photon emission,
called paired superradiance
(PSR for short), $|e\rangle \rightarrow |g\rangle + \gamma + \gamma$, 
becomes easily detectable \cite{macro-coherence}, its rate $\propto n^2 V$,
with $V$ a macroscopic target volume,
unlike the case of usual SR
limited by $V \propto$ the photon wavelength squared.
PSR has a distinct signature:
two photons are back to back emitted and
have exactly the same energy.

We propose in this work to use for the target of RNPE
a coherent state of atoms interacting with static field condensate
(we call this as condensate for simplicity).
The condensate is a limiting case of multiple soliton
solutions, as presented below.
Both solitons and condensate are proved stable
against PSR, but unstable for RNPE.

PSR, emitting a highly correlated pair of two photons, 
is interesting from points of
application such as quantum entanglement.
Artificial destruction of solitons and condensate, 
which can be easily realized by
a sudden application of electric pulse (thus 
abruptly changing the
dielectric constant),  provides
the most efficient mechanism of PSR emission
yet to be discovered.
If we successfully destroy solitons for PSR under complete control,
solitons may become qubits for quantum computing.

On the other hand,
creation and subsequent long time control
of the condensate
removes the most serious PSR background for RNPE.
We compute  macro-coherent RNPE rate $\propto n^2 V$ of
condensate decay and study sensitivity of spectral rates 
(spectral shape and event rate) to
parameters of the neutrino mass matrix, most importantly
the fundamental parameter of CP violating Majorana phases;
the parameter of central importance in
explaining the matter-antimatter imbalance
of the universe.
RNPE spectrum shape from the condensate decay is 
time independent after condensate formation and 
the most unambiguous tool for this process.

Our method uses laser to trigger
RNPE at non-resonant frequencies,
which should be a great merit since the trigger
is not destructive to target atoms.

The natural unit $\hbar = c= 1$ is used in formulas of this paper.

{\bf Effective atomic Hamiltonian and Maxwell-Bloch equation}
\hspace{0.2cm}
We consider atoms that consist of three levels of energies
$\epsilon_g < \epsilon_e < \epsilon_p$.
The state $|e\rangle $,
for example $^1$D$_2$-state of Ba
low lying levels, is forbidden to decay to $|g\rangle $ by E1 transition, 
while E1 transitions from
$|p \rangle$ to $|e \rangle$ and $|g \rangle$
may both be allowed. 
The important part of Hamiltonian is 
derived  \cite{narducci}, \cite{my-10-10}
by eliminating time memory effects of $|p\rangle$, 
$^1P_1$ in Ba,
and by making a slowly varying envelope approximation
of one field mode  propagating in a direction.
The resulting effective Hamiltonian is
restricted to two levels, $|e \rangle$ and $|g \rangle$,
interacting with field $E$ of frequency $\omega$
and a definite polarization.
The $2\times 2$ matrix elements $\mu_{ab}\,,a,b= e,g,$ are Stark energies;
a product of two dipole (E1 or M1) transition elements to
$|p\rangle$  times the electric field squared.
Dipole transition elements are
related to measurable  decay rates 
$\gamma_{pa}\,, a=e, g$ from $|p\rangle$ to $|a\rangle$, thus
\begin{eqnarray}
&&
\mu_{aa} = \frac{6\pi \gamma_{pa}}{\epsilon_{pa}^2 (\epsilon_{pa}^2-\omega^2)}
\,, \hspace{0.5cm} (a=e\,,  g)
\,,
\\ &&
\mu_{eg}=\mu_{ge} = \frac{3\pi (\epsilon_{pe}+\epsilon_{pg})}{2
 (\epsilon_{pe}-\omega)(\epsilon_{pg}+\omega)}
\sqrt{\frac{\gamma_{pe}\gamma_{pg}}{\epsilon_{pe}^{3}\epsilon_{pg}^{3}}}
\,.
\end{eqnarray}
We ignored the spin multiplicity factor $2J_a+1$ in
the relation $d_{ab}^2$ to $\gamma_{ab}$.
The final PSR rate formula should be multiplied
by $(2J_p+1)/(2J_e+1)$ if one includes this multiplicity.

The  equation for the polarization vector $\vec{R}$
(3 bilinears of amplitudes times the target number density $n$),
called the Bloch equation, is derived from the Schr\"{o}dinger
equation, and may be written as 
\(\:
\partial_t \vec{R} = |E|^2{\cal M} \vec{R} \,,
\:\)
where elements of
$3\times 3$ anti-symmetric matrix ${\cal M}$ are linear
combinations of $\mu_{ab}$.
When this equation is combined with the Maxwell
equation, written as 
\(\:
(\partial_t + \partial_x )|E|^2 = \omega \mu_{ge}|E|^2 R
\,,
\:\)
with $R$  a component of $\vec{R}$,
a closed set of equations follows, to describe 
spacetime evolution of polarization 
and propagating field  \cite{narducci}, \cite{my-10-10}.

When relaxation processes are ignored,
one can introduce the tipping angle $\theta(x,t)$
by $R(x,t) = n \cos \theta(x,t)$.
The Bloch equation is then reduced to a relation of $\theta(x,t)$ to
the electric field strength; $|E(x,t)|^2 = \partial_t \theta/\mu$
with $\mu= \sqrt{(\mu_{ee}-\mu_{gg})^2 + 4\mu_{ge}^2}/4$.
The field $\theta(x,t)$ is an analogue of
the area for field propagation in the two-level problem
\cite{coherent light propagation in 2 level}.
The Maxwell equation in terms of $\theta(x,t)$ is
\begin{eqnarray}
&&
(\partial_t + \partial_x)\theta = \alpha_m (- \cos \theta + A)
\,,
\label{angle eq}
\\ &&
\alpha_m = 6\pi\sqrt{\frac{\gamma_{pe}\gamma_{pg}}{\epsilon_{pe}^3 \epsilon_{pg}^3}}
\frac{\omega n}{\epsilon_{pe}+ \epsilon_{pg}}
\,,
\end{eqnarray}
where $\epsilon_{ba} = \epsilon_{b} - \epsilon_a$ is the atomic energy difference.
For the Ba D-state, 
$\alpha_m \sim 2.4 \times 10^{-6}$cm$^{-1} (n/10^{12}{\rm cm}^{-3})$
at $\omega = \epsilon_{eg}/2$.
Both $\alpha_m$ and $ \mu$ depend on $\omega$.
The  non-linear equation (\ref{angle eq}) describes dynamics 
of a fictitious pendulum  under 
friction periodically varying $\propto \alpha_m \sin \theta$
at its location $\theta$.

For $|A|\leq 1$, the tipping angle is restricted
to a finite $\theta-$region of $\leq 2\pi$.
The propagation problem in this case has been analytically solved
in \cite{my-10-10} in terms of arbitrary initial data.
Hence the system appears integrable in the mathematical sense.
Typical solutions describe
multiple splitting of pulses and
their compression when they propagate in a long coherent medium,
as fully explained in \cite{my-10-10}.
The number of split pulses is given by
the initial pulse area $\theta(-\infty,\infty)$ divided by $2\pi$.
This behavior of pulse in medium is a symptom of instability,
and pulses stabilize via PSR.
It is thus anticipated that stable
objects against PSR exist; solitons.

{\bf Soliton solutions}
\hspace{0.2cm}
There are two types of analytic solutions for solitons;
$ |A|=1$  giving a single soliton of quantized area $2\pi$
and $|A|>1$ the multiple soliton.
The case $|A|< 1$ is unphysical since 
an excited state of population $n \cos \theta \neq - n$ exists
at $\xi = \pm \infty$.
The case of $ |A|=1$ solution of area $2\pi$ has been
obtained  in \cite{my-10-10} by using a different method.

We look for soliton solutions by assuming 
one variable dependence of $ x - vt$
for a soliton of velocity $v$ and
by reducing the partial differential equation to
an ordinary one.
The solution for $A=1$ thus obtained has a Lorentzian shape of flux
and the population given by
\begin{eqnarray}
&&
|E_s(x,t)|^2 = \frac{2\alpha_m }{\mu}
\frac{v(1-v)}{\alpha_m^2 (x-vt)^2 + (1-v)^2}
\,,
\\ &&
\cos \theta_s (x,t) = \frac{-\alpha_m^2 (x-vt)^2 + (1-v)^2}{\alpha_m^2 (x-vt)^2 + (1-v)^2}
\,.
\end{eqnarray}
The soliton size is $O[1/\alpha_m]$, and
its field flux is of order, 
$\alpha_m/\mu \sim 30 {\rm W mm}^{-2} (n/10^{18}{\rm cm}^{-3})$
for the Ba soliton at $\omega=\epsilon_{eg}/2$.

This method applied to the $|A| > 1$ case, on the other hand, gives
a new class of solutions given by
\begin{eqnarray}
&&
|E(x,t)|^2 = \frac{\alpha_m }{\mu}\frac{v}{1-v}
\frac{A^2-1} {A - \cos X }
\,, \hspace{0.5cm}
\cos \theta(x,t) = \frac{A \cos X -1}{A - \cos X}
\,,
\label{multiple soliton}
\\ &&
X = \frac{\alpha_m \sqrt{A^2-1}}{1-v}(x-vt)
\,.
\end{eqnarray}
Unlike the single peak for $|A|=1$,
the field flux given by (\ref{multiple soliton}) has infinitely many peaks equally spaced,
describing multiple soliton solutions in medium.

For a finite length of medium one may
impose the boundary condition of
no excited state at two target ends of $x= \pm L/2$.
This gives a condition,
$\alpha_m L \sqrt{A^2-1}/(1-v) = 2\pi (2n_s-1)\,, n_s= 1,2,\cdots$.
The quantity $\sqrt{A^2-1}$ is  of order, and the soliton
number density $\sim n_s/(\alpha_m L)$.

Solitons may both emit and absorb photons within medium,
their rate difference $\propto \cos \theta |E(x,t)|^2$.
This quantity, when integrated in the entire medium supporting a soliton,
gives an integral of a total derivative $\propto \partial_x  \sin \theta$,
hence vanishes for the quantized area of $\Delta \theta = 2\pi$.
This proves the soliton stability against PSR.

{\bf Field condensate}
\hspace{0.2cm}
One may consider the limit of large soliton density, 
$n_s/\alpha_m L \sim \sqrt{A^2-1}/4\pi \rightarrow \infty$, simultaneous with
the limit $v \rightarrow 0$.
Denoting  $A = \sqrt{(\eta/v)^2 + 1}$ with $\eta$ kept constant,
one has 
\begin{eqnarray}
&&
|E_c(x.t)|^2 = 
\frac{\alpha_m \eta^2}{\mu\left(\eta - v\cos (\alpha_m \eta (x-vt)/v)\right)}
\approx \frac{\eta \alpha_m}{\mu}
\,,
\end{eqnarray}
thus an almost constant field flux is derived.
The population $\propto \cos \theta$ oscillates, with the
time period $\tau = 2\pi /(\alpha_m \eta)$ and the
space period $\tau v$.
The parameter $\eta$  is $4\pi \times$ soliton density $\times$ 
soliton velocity. 

Practically,
the shortest spatial period is limited by the inter-atomic
distance $d$.
By identifying the period $\tau v$ with $d$, 
one finds $\eta \sim 2\pi v/(\alpha_m d)$,
hence $\tau = d/v$. 
As $v \rightarrow 0$, $\tau \rightarrow \infty$, and
the target becomes fully excited  with 
$\cos \theta = 1$.
For the Ba $^1$D$_2$-state, the relevant numerical value is
\(\:
\alpha_m d \sim 4\times 10^{-7}(n/10^{18}{\rm cm}^{-3})^{2/3}
\,.
\:\)

The limit taken here gives a constant field $\eta \alpha_m/\mu$
and the full excitation of target everywhere
(strictly, this is true for an infinitely long medium).
This is the state of field condensate we use for RNPE.
Field condensate can be created by trigger
laser irradiation from multiple directions,
since it has no memory of a particular direction.

The  stability analysis around
the condensate  can be made, taking $E=E_c + \delta E\,, \theta=\theta_c+
\delta \theta$ with
$E_c\,, \theta_c=0$ the condensate solution.
By keeping linear terms $\propto \delta E, \delta \theta$ in
the Maxwell-Bloch equation, with
$\delta E, \delta \theta \propto e^{-i\omega t}$
for time dependence, the perturbation equation 
\(\:
\partial_x \delta E = i (\omega + \alpha_m^2 \eta/\omega)\delta E 
\:\)
gives a bounded and purely oscillatory solution,
indicating the stability of field condensate.

{\bf PSR rate at soliton and condensate destruction}
\hspace{0.2cm}
We first mention PSR rate without soliton
creation. The
PSR rate without trigger is
$ \mu_{ge}^2 \epsilon_{eg}^4n^2V/(2^9 \pi^2)$, 
which is numerically 
$\sim 0.5 {\rm MHz}(n/10^{12}{\rm cm}^{-3})^2 V/{\rm cm}^{3}$
for Ba.
Under a strong trigger of flux $|E|^2$,
the rate for a target of length $L$ becomes \cite{my-10-10}
\begin{eqnarray}
&&
\frac{\mu_{ge}^2 \epsilon_{eg}n^2VL|E|^2}{32 \pi}
\,.
\label{psr from field}
\end{eqnarray}
Although the rate for $|E|^2 \approx 10^6$Wcm$^{-2}$ is large,
time structure of PSR
is complicated \cite{my-10-10}.

PSR after soliton formation
occurs only at its destruction, without absorption
from $|g\rangle$.
The emission rate from $|e\rangle$
is $\propto (1+ \cos \theta)|E_s|^2/2$.
One may compute rates based on perturbative methods, 
in which one of the photons belongs to the soliton pulse.
The other photon is emitted backward to the soliton 
propagation direction, with exactly the same energy.
The large rate enhancement $\propto n^2 V$ is understood by
the momentum conservation among
emitted particles, 
implying $e^{i(\vec{k} + \vec{k'})\cdot \vec{x}} = 1$.

The PSR rate at soliton destruction is (taking $L = dx$ in eq.
(\ref{psr from field}) )
\begin{eqnarray}
&&
d\Gamma(x,t) = \frac{\mu_{ge}^2 \alpha_m \epsilon_{eg} n^2V }{16\pi \mu}
\frac{v(1-v)}{\alpha_m^2 (x-vt)^2 + (1-v)^2}
dx
\,.
\end{eqnarray}
The rate remains large during a time of
\begin{eqnarray}
&&
\Delta t = \frac{1-v}{ v\alpha_m }
\sim 14 \mu {\rm sec}\frac{1-v}{v}\frac{10^{12}{\rm cm}^{-3}}{n}
\,.
\end{eqnarray}
The space integrated rate per soliton is
\begin{eqnarray}
&&
\frac{\mu_{ge}^2  \epsilon_{eg}  v n^2 V}{16 \mu}
\sim
5 \times 10^{15} {\rm Hz} \frac{v n^2 V}{10^{24}{\rm cm}^{-3}}
\,,
\end{eqnarray}
(numbers for Ba)
a formula valid for a target of length $L \gg 1/\alpha_m$.
For a short target of $L \leq 1/\alpha_m$
the rate is reduced by $\alpha_m L/(\pi (1-v)\,)$.
The integrated rate for long
target is by many orders ($\sim 10^{8}$) larger than
the trigger-less PSR rate.
The prolonged time of $O[1] \mu$sec and its simple
profile structure has a number of merits
of easier PSR identification such as the back to back
two photon coincidence measurement.
PSR rate at condensate destruction is larger by
$\eta\alpha_mL/(v\pi)$ than at the single soliton destruction.

{\bf Effect of relaxation}
\hspace{0.2cm}
There are a number of processes
that might destroy coherence.
One of them is given by a field decay, introduced by
a term $\kappa |E|^2$ in the Maxwell equation.
This modifies the basic equation (\ref{angle eq}) by
an additional term $-\kappa \theta$.
With the ansatz of variable 
dependence of $\xi =x-vt$, this equation is
\begin{eqnarray}
&&
(1-v)
\frac{d \theta }{d \xi} = - \alpha_m  \cos \theta -  \kappa  \theta
\,.
\label{basic eq 2}
\end{eqnarray}

Direct numerical integration of eq.(\ref{basic eq 2})
gives distorted quasi-soliton solutions.
Their profile, although distorted, is unchanged as they propagate.
Quasi-solitons exist only for $\kappa < \kappa_c$ where 
$ \kappa_c \sim 0.725 \alpha_m$,
indicating a threshold of dissipation.
Calculation gives the PSR rate $=$ (PSR rate at pure soliton destruction) 
$\times \Delta (\theta + \sin \theta)/2\pi$
(the difference $\Delta$ to be taken at two target ends).
This rate is smaller than the one without dissipation, but
not very much less, unless $\kappa$ is very close to the threshold $\kappa_c$.
The condition of a sizable PSR rate, 
the relaxation constant $\kappa < O[\alpha_m]$,
implies that $\kappa < O[0.07]{\rm MHz}(n/10^{12}{\rm cm}^{-3})$ for the Ba target.

{\bf RNPE}
\hspace{0.2cm}
The effective Hamiltonian for RNPE,
\begin{eqnarray}
&&
\frac{G_F\vec{S}_e\cdot\sum_{ij}c_{ij}\nu_j^{\dagger}\vec{\sigma}\nu_i}{\epsilon_{pg} -\omega}
 \vec{d}\cdot\vec{E}
 \,,
\end{eqnarray}
gives the amplitude for a single atom, where 
$\vec{S}_e$ and $\vec{d} $ are electronic spin and dipole operators.
To give  large matrix elements for these, 
we consider deexcitation of $|e\rangle$
of the angular momentum $J=2$ or $J=0$ to  $|g\rangle$
of $J=0$ via  $|p\rangle$
of $J=1$, realized by
rare gas and alkhali earth atoms.
Six measurable  constants, $c_{ij}$'s,
given by $ c_{ij}=U_{ei}^*U_{ej} - \delta_{ij}/2$
with $U$ the unitary matrix relating the neutrino flavor to the mass eigenstate,
contain mixing angles and 
Majorana CP phases \cite{my-06}, \cite{pv}.

The field operator $\nu_i$ for the Majorana neutrino is a superposition of
 annihilation ($b_i$) and creation ($b_i^{\dagger}$) operator of the same
Majorana particle, while for the Dirac neutrino
it is a sum of two distinct operators; particle annihilation ($b_i$) 
and anti-particle creation ($d_i^{\dagger}$).
Thus, the $\nu_i \nu_j$ pair emission amplitude for $i\neq j$
has the form, 
\( \:
b_i^{\dagger}b_j^{\dagger}(c_{ij} - c_{ji})/\sqrt{2}
=i \sqrt{2}\Im c_{ij}b_i^{\dagger}b_j^{\dagger}
\:\)
for the Majorana case,
and $c_{ij}b_i^{\dagger}d_j^{\dagger}$ for the Dirac case.
Condensate RNPE decay rate of field $|E_c|^2 \sim \eta \alpha_m/\mu$
is a sum of 6 $\nu_i \nu_j$ pair emission;
\begin{eqnarray}
&&
\frac{48 n^2 V \eta\alpha_m G_F^2 \gamma_{pg} }
{\epsilon_{pg}^3(\epsilon_{pg}-\omega)^2 \mu}
\sum_{ij}B_{ij}I_{ij}(\omega)
\,.
\label{rnpe rate}
\end{eqnarray}
For $i\neq j$,
$B_{ij}=  (\Im c_{ij})^2$ for the Majorana case and
$B_{ij}=  |c_{ij}|^2$ for the Dirac case,
while $B_{ii}=  |c_{ii}|^2$ for both cases.
Factors, $\alpha_m$ and $\mu$,
attributed to condensate parameters, 
involve intermediate $|p\rangle$.
The state  $|p\rangle$ that
gives the largest condensate factor may be different
from the intermediate state that gives the largest RNPE rate.
In the Yb case, $|p \rangle = 6s6p ^3P_1$ for the largest condensate factor
and $|p \rangle = 6s6p ^1P_1$ for the largest RNPE rate.

\begin{figure*}[htbp]
 \begin{center}
 \epsfxsize=0.5\textwidth
 \centerline{\epsfbox{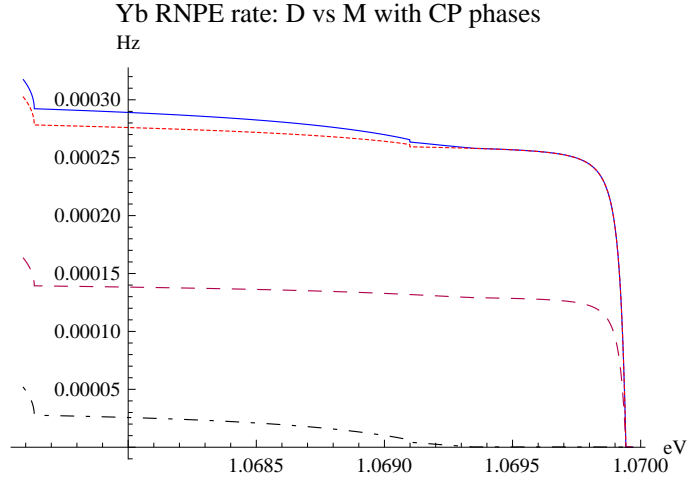}} \hspace*{\fill}
   \caption{Yb RNPE photon energy spectrum in
(11) $\sim$ (33) region from condensate decay.
Dirac case and 3 Majorana cases of different $(\alpha, \beta)$
are plotted; Dirac  in blue,
($\pi/2,0$) Majorana in dotted or short dashed red,
($0, \pi/2$) Majorana in broken black,
and ($\pi/4, -\pi/4$) Majorana in dashed purple.
Neutrino masses of $(m_1, m_2, m_3) = (50,10,1)$meV, and
 cosine angles $1/\sqrt{2},\sqrt{3}/2,\sqrt{0.97}$ are assumed for all.
 The Majorana pair emission rate of $(\alpha, \beta)=$ (0,0) 
below (3,3) neutrino threshold is
by $\sim 10^{-3}$ smaller than the Dirac rate for these masses.
Assumed target parameters are
$n=10^{21}$cm$^{-3}$, $V=1{\rm cm}^3$, 
and $\eta = 10^3$.
   }
   \label{Yb RNPE 2}
 \end{center} 
\end{figure*}

The function $I_{ij}(\omega)$ in the formula (\ref{rnpe rate}) is
given by an energy integral
arising from the two neutrino phase space.
The integral in a symmetric form is given in terms of neutrino
energies $E_i\,, i=1,2$,
\begin{eqnarray}
&&
I_{ij}(\omega) = \omega
\int_{0}^{\infty} \int_{0}^{\infty} dE_1 dE_2 \delta(E_1+E_2 + \omega - \epsilon_{eg})
\theta(C_{ij}(E_1,E_2))
\left(\frac{K_{ij}^{(1)}}{12} + \frac{3K_{ij}^{(2)}}{4} \right)
\,,
\\ &&
K_{ij}^{(1)} = 2 G_{ij}^{(1)}(E_1,E_2)
\,,\;
K_{ij}^{(2)} = - G_{ij}^{(1)}(E_1,E_2)+G_{ij}^{(2)}(E_1,E_2) 
+ \frac{E_1E_2 - \delta_M m_i m_j}{\omega^2}
\,,
\\ &&
G_{ij}^{(1)}= \frac{1}{8}
 + \frac{E_1^2 + E_2^2 - m_i^2-m_j^2}{4\omega^2}
-3\frac{(E_1^2 - E_2^2 - m_i^2+m_j^2)^2}{8\omega^4}
\,, 
\\ &&
G_{ij}^{(2)}= \frac{1}{8} -\frac{E_1^2 + E_2^2 - m_i^2-m_j^2}{4\omega^2} 
+ \frac{(E_1^2 - E_2^2 - m_i^2+m_j^2)^2}{8\omega^4}
\,,
\end{eqnarray}
where the boundary region is given by $C_{ij}(E_1,E_2) \geq 0$
with
\begin{eqnarray}
&&
C_{ij}(E_1,E_2) = 
(E_1^2 + E_2^2 - m_i^2-m_j^2 -\omega^2)^2
- 4(E_1^2 - m_i^2)(E_2^2 - m_j^2)
\,,
\end{eqnarray}
and $\delta_M = 1$ for the Majorana and $\delta_M = 0$ for the Dirac case.
In this calculation,
averaged electron spin matrix elements,
$\langle (\vec{k}\cdot\vec{S}_e/\omega)^2 \rangle = 1/12\,,
\langle \vec{S}_e^2 \rangle = 3/4$, are used.
The resulting spectrum given by
(\ref{rnpe rate}) sharply rises at each $(ij)$ threshold, 
a feature characteristic of 3 particle
emission of massless $\gamma$ and nearly massless $\nu_i, \nu_j$, when both 
the momentum and the energy conservation hold.

The limiting case of 3 massless neutrinos gives
RNPE rate of the condensate decay,
\begin{eqnarray}
&&
\frac{
G_F^2 \gamma_{pg}\epsilon_{eg}^2 n^2 V \eta \alpha_m }{
\mu \epsilon_{pg}^5}f(\frac{2\omega}{\epsilon_{eg}})
\,, \hspace{0.5cm}
f(x) = \frac{9(324-540x + 245x^2)}{32(1- \epsilon_{eg}x/(2\epsilon_{pg})\,)^2}
\,.
\end{eqnarray}
The coefficient in front of the function $f(2\omega/\epsilon_{eg})$ is
not rapidly varying with the photon energy 
$\omega $ in the neutrino threshold regions,
and on the average over the photon energy to $0.95 $eV,
$\sim 8 \times 10^{-5}$Hz for Yb of
$n= 10^{21}{\rm cm}^{-3}\,, V = 1{\rm cm}3\,, \eta = 10^3$.

Experiments for the neutrino
spectroscopy are to be performed keeping the
macro-coherence of the condensate.
The initial trigger frequency $\omega \leq \omega_{11}$ for RNPE of
$\omega_{11}= \epsilon_{eg}/2 - 2 m_1^2/\epsilon_{eg}$,
with $m_1$ the smallest neutrino mass,
is reset each time for measurements of rate
and parity violating quantities \cite{pv} at different
$\gamma$ energies of the continuous spectrum.
The energy resolution of RNPE spectrum is thus
determined by the precision of trigger
frequency $\omega$, and not by detected photon energy.
This is a key element for successful
implementation of the precision  neutrino
mass spectroscopy, which must resolve photon energies at 
the $\mu$eV level or less, since
the $(ij)$ threshold rise at  
$\omega_{ij} = \epsilon_{eg}/2 - (m_i+m_j)^2/(2\epsilon_{eg})$
is separated only a little from the half energy $\epsilon_{eg}/2$ of 
dangerous PSR.

Calculated rates are sensitive to
Majorana CP phases $\alpha, \beta$ defined by
\(\:
U_{e2} \propto e^{i\alpha}
\,,
U_{e3} \propto e^{i\beta}
\,.
\:\)
Rate rises at $(12),(13), (23)$ thresholds are 
$\propto (|c_{12}|^2\sin^2 \alpha\,,|c_{13}|^2\sin^2 \beta\,, |c_{23}|^2\sin^2(\alpha-\beta)\,)$
(the Dirac case given by $|c_{ij}|^2$ without $\alpha, \beta$ phase), 
to be further multiplied by an integrated Majorana interference factor of
$(1- m_i m_j/E_1E_2)$ with $E_i$ neutrino energies.
For example,
4 cases of $(\alpha,\beta) = (0,0),(\pi/2,0), (0,\pi/2), (\pi/4,-\pi/4)$
give 3 large threshold phase factors of
$(\sin^2 \alpha, \sin^2\beta, \sin^2 (\alpha-\beta)\,)
= (0,0,0),(1,0,1),(0,1,1),(1/2,1/2,1)$
at $(12),(13), (23)$.
Since $|c_{12}|^2 \approx 3 \times (|c_{11}|^2, |c_{22}|^2)$
under the given oscillation parameters,
the magnitude $\sin^2 \alpha $ is most important in
rate comparison at low threshold regions, and a small $\sin^2 \alpha $
gives a much smaller Majorana rate than the Dirac rate
below (33) threshold.
We know of no other measurable quantity of this
high sensitivity to $\alpha$ and $\beta$.
Within a given range of neutrino parameters,
the easiest observable might be the Majorana phase,
as illustrated in our figures.
Our proposed experiment is not sensitive to
the other CP phase $\delta$, which however may be
determined by future neutrino oscillation experiments.
Determination of all low energy phases, $\alpha, \beta, \delta$,
is a requisite for a better understanding of leptogenesis
\cite{davidson-ibarra}.

\begin{figure*}[htbp]
 \begin{center}
 \epsfxsize=0.5\textwidth
 \centerline{\epsfbox{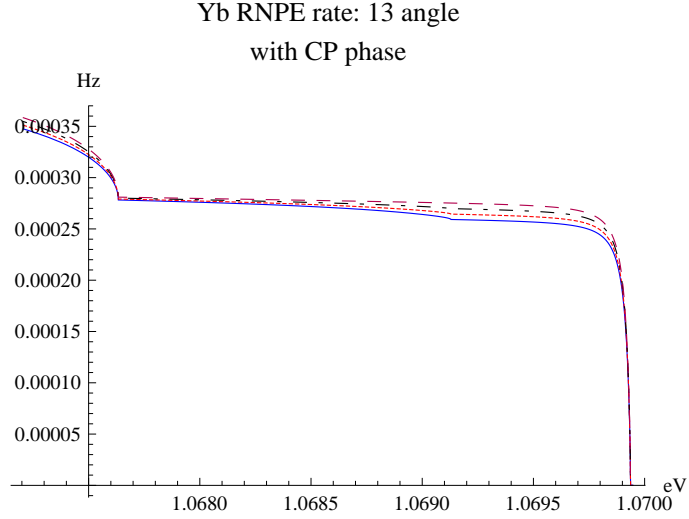}} \hspace*{\fill}
   \caption{Yb RNPE photon energy spectrum for different
   $\sin^2 \theta_{13}$ values, 0.03 in blue, 0.02 in dotted red, 
   0.01 in broken black, and 0 in dashed purple, all for $(\alpha, \beta)=(\pi/2,0)$.
Neutrino masses of $(m_3,m_2,m_1) = (50,10,5)$meV, and
 cosine angles $1/\sqrt{2},\sqrt{3}/2$ 
for $\cos \theta_{23},\cos \theta_{12}$ are assumed.
Assumed target parameters are
$n=10^{21}$cm$^{-3}$, $V=1{\rm cm}^3$, 
and $\eta = 10^3$.
   }
   \label{Yb RNPE 13}
 \end{center} 
\end{figure*}

\begin{figure*}[htbp]
 \begin{center}
 \epsfxsize=0.5\textwidth
 \centerline{\epsfbox{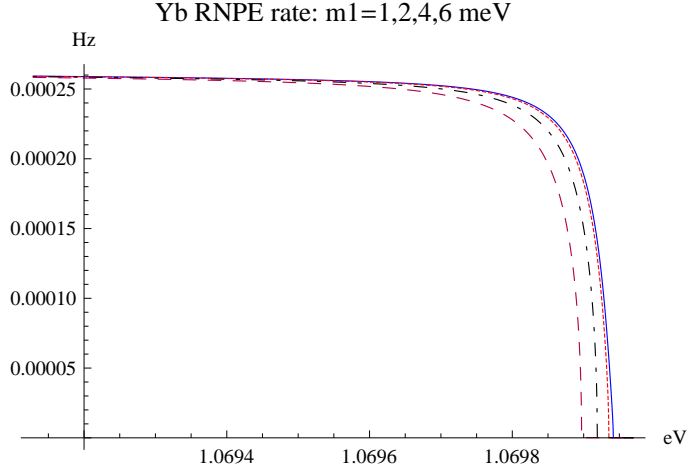}} \hspace*{\fill}
   \caption{Yb RNPE photon energy spectrum for different
   $m_1$ values; 1 meV in blue, 2 meV in dotted red, 
   4 meV in broken black, and 6 meV in dashed purple,
all for $(\alpha, \beta)=(\pi/2,0)$.
Other neutrino masses are constrained by neutrino oscillation experiments,
and cosine angles $1/\sqrt{2},\sqrt{3}/2,\sqrt{0.97}$ 
for $\cos \theta_{23},\cos \theta_{12}, \cos \theta_{13}$ 
are assumed.
Assumed target parameters are
$n=10^{21}$cm$^{-3}$, $V=1{\rm cm}^3$, 
and $\eta = 10^3$.
   }
   \label{Yb RNPE mass1}
 \end{center} 
\end{figure*}

\begin{figure*}[htbp]
 \begin{center}
 \epsfxsize=0.5\textwidth
  \centerline{\epsfbox{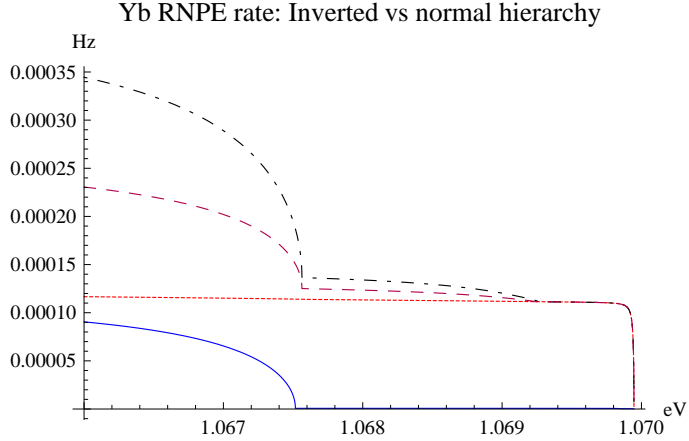}} \hspace*{\fill}
   \caption{Yb RNPE rate; case of normal and
inverted mass hierarchies for different $(\alpha,\beta)$ values;
   normal (0,0)  (blue), 
inverted (0,0) (dotted red), inverted ($\pi/2$,0) (broken black),
   inverted ($\pi/4, -\pi/4$) (dashed purple), 
 assuming $m_1 = 5$ meV (other masses constrained
by oscillation data) and the same
mixing as Fig(\ref{Yb RNPE mass1}).
$ n=10^{21}{\rm cm}^{-3}, V=1{\rm cm}^{3}, \eta = 10^3$.
}
   \label{Yb rnpe rate vs inverted mass}
 \end{center} 
\end{figure*}

Distinction of Majorana and Dirac neutrinos is
possible by the interference
effect of identical Majorana fermions \cite{my-06},
giving different  rates in the vicinity of thresholds.
Rate difference of Majorana and Dirac pair emission
 is larger for larger Majorana CP phases,
as illustrated in Fig(\ref{Yb RNPE 2}).
Experimentally, the spectral rate is fitted under an assumption
of Majorana or Dirac neutrino and
either hypothesis is verified by a good quality of fitting.

One possible serious background against RNPE might be the
trigger-less SR
due to the achieved excellent coherence.
This process has a monochromatic
spectrum at $\epsilon_{eg}/2$ different from RNPE, nevertheless it might 
become dangerous, destroying the initial state.
This can be avoided by choosing $J=0 \rightarrow 0$ transition,
which forbids single photon emission, complete to any order,
hence SR altogether.
Alkhali earth atoms have level structure of this 
angular momentum configuration.
Yb and Hg atoms have levels of a similar nature,
giving state candidates of
$|e\rangle = (6s 6p)^3P_0\,, |g\rangle = (6s^2)^1S_0\,,
|p \rangle = (6s 6p)^1P_1$.
Incidentally,
two photons emitted by $0 \rightarrow 0$ RSR
have perfectly correlated polarizations, and
may serve as an excellent device of quantum 
entanglement.

The calculated Yb $0 \rightarrow 0$ RNPE rate for $(\alpha, \beta) = (0,0)$
averaged over all photon energies is $\sim 3 \times 10^{-4}$Hz
for $n=10^{21}{\rm cm}^{-3}\,, V= 1 {\rm cm}^{3}\,, \eta = 10^3$
(a factor to be better understood) and is by $\sim 70$
larger than the corresponding Xe $2 \rightarrow 0$ rate.
When the Yb experiment at each photon energy $\omega$ lasts for a day, 
its event number  becomes
$O[30] $ if $\omega$ is in the energy range of Fig(\ref{Yb RNPE 2}).
This event number is further increased by $f$ if
one repeats condensate formation  with a
cycle time of $1/f$ sec.
We show in Fig(\ref{Yb RNPE 2})  and Fig(\ref{Yb RNPE 13}) 
the spectral rate for various combinations of
CP phases and the mixing angle $\theta_{13}$.
Sensitivity to neutrino masses, in particular to $m_1$ values,
is shown in Fig(\ref{Yb RNPE mass1}).
Determination of $m_1$ of a few meV range
requires a high statistic data near (11) threshold.
Distinction of normal 
(3 neutrino masses given by $m_1, m_2=\sqrt{0.01^2+ m_1^2}, m_3 = \sqrt{0.05^2
+ 0.01^2+ m_1^2}$ eV's) 
vs inverted 
(3 masses of $m_1, m_2=\sqrt{0.05^2+ m_1^2}, m_3 = \sqrt{0.05^2
+ 0.01^2+ m_1^2}$ eV's) hierarchies is
most dramatic, 
as seen in Fig(\ref{Yb rnpe rate vs inverted mass}),
hence its determination is easier.

RNPE rate of condensate decay
increases like $\propto n^3$, effective with $\alpha_m \propto n$,
as the density $n$ increases. 
The event number from a single soliton decay is smaller
than from the condensate decay,
typically by $1/(\eta \alpha_m L)$ for a target length
$L\gg 1/\alpha_m$ that contains a soliton.

In an ideally coherent medium,
field condensate never
emits PSR.
In practice, there may be a variety of
environmental effects that cause a leakage PSR,
a potential background to RNPE.
One of these effects is a random fluctuation
of dielectric constant, most
simply due to a density fluctuation 
$\sqrt{\delta n^2} $. 
The resulting leakage PSR rate is estimated as
\begin{eqnarray}
&&
\frac{
3 \mu_{ge}^2 \epsilon_{eg}n \sqrt{\delta n^2} V \alpha_m L
e^{-(\omega - \epsilon_{eg}/2)^2/\Delta_m^2}}{32\pi \mu} \times
 \frac{d\omega}{\sqrt{\pi }\Delta_m}
\,,
\end{eqnarray}
for a target length $L$.
We used a Gaussian frequency distribution of width $\Delta_m$
for the trigger.
We have computed the Yb leakage PSR rate using
$\sqrt{\delta n^2}/n=$5\%  
and   $\Delta_m =1$GHz.
The calculated Yb RNPE near (12) threshold of  $m_1 =2$meV is found much larger 
than the background PSR.
The leakage PSR  becomes larger than RNPE, only at 
photon energies $\leq 2 \mu$eV away from the first (11) threshold.

In summary,
our proposed method of precision neutrino mass
spectroscopy is most sensitive to 
Majorana/Dirac distinction and to
$\alpha, \beta$ measurements.
It is worthwhile to experimentally investigate both
formation and long time control of solitons and
condensate, which is of crucial importance
for controlled detection of PSR and RNPE.
Some rudimentary method of efficient soliton
formation has been suggested in \cite{my-10-10}.

\vspace{0.3cm}
{\bf Acknowledgements }
\hspace{0.2cm}
I should like to thank N. Sasao and
members of SPAN collaboration for discussion on
experimental aspects of this subject,
and M. Tanaka for discussion on an aspect of leptogenesis.

This research was partially supported by Grant-in-Aid for Scientific
Research on Innovative Areas "Extreme quantum world opened up by atoms"
(21104002)
from the Ministry of Education, Culture, Sports, Science, and Technology.

\end{document}